\documentclass[12pt]{article}

\DeclareSymbolFont{bletters}{OML}{cmm}{bx}{it}
\DeclareMathSymbol{\bdl}{\mathord}{bletters}{"0E}

\begin{document}

\title {
$${}$$
{\bf Functional integration}\\
{\bf with ``automorphic'' boundary conditions}\\
{\bf and correlators of z-components of spins}\\
{\bf in the $XY$ and $XX$ Heisenberg chains}
$${}$$}

\author{C. MALYSHEV\\
$${}$$
{\it V. A. Steklov Institute of Mathematics,}\\
{\it St.-Petersburg Department,}\\
{\it Fontanka 27, St.-Petersburg, 191023, RUSSIA}\\
E-mail: malyshev@pdmi.ras.ru}

\maketitle

\def \al{\alpha}
\def \be{\beta}
\def \ga{\gamma}
\def \dl{\delta}
\def \ep{\varepsilon}
\def \ze{\zeta}
\def \th{\theta}
\def \ka{\varkappa}
\def \la{\lambda}
\def \si{\sigma}
\def \ph{\varphi}
\def \om{\omega}
\def \Ga{\Gamma}
\def \Dl{\Delta}
\def \La{\Lambda}
\def \Si{\Sigma}
\def \Ph{\Phi}
\def \Om{\Omega}
\def \cA{\cal A}
\def \cB{\cal B}
\def \cC{\cal C}
\def \cD{\cal D}
\def \cN{\cal N}
\def \BC{I\!\!\!\! C}
\def \BD{I\!\!\!\! D}
\def \BZ{Z\!\!\!Z}
\def \IM{\Im}
\def \RE{\Re}
\def \1{^{-1}}
\def \cd{\partial}
\def \CD{{\cal D}}
\def \CA{{\cal A}}
\def \CM{{\cal M}}
\def \CL{{\cal L}}
\def \at{{\rm arctan}\,}
\def \ch{{\rm ch}\,}
\def \sh{{\rm sh}\,}
\def \th{{\rm th}\,}
\def \ld{\ldots}
\def \CU{{\cal U}}
\def \BQ{I\!\!Q\!\!I}
\def \bQ{{\bf Q}}
\def \vt{\vartheta}
\def \w{\widetilde}
\def \h{\widehat}
\def \d{\dagger}
\def \t{\times}
\def \l{\langle}
\def \r{\rangle}
\def \Tr{{\rm Tr}\,}
\def \tr{{\rm tr}\,}
\def \diag{{\rm diag}\,}
\def \Det{{\rm Det}\,}
\def \z{\zeta}

\begin{abstract}
\noindent
Representations for the generating functionals of 
static correlators of $z$-components of spins in 
the $XY$ and $XX$ Heisenberg spin chains are obtained 
in the form of sums of the fermionic functional 
integrals. The peculiarity of the functional integrals 
in question is because of the fact that the integration 
variables depend on the imaginary time 
``automorphically''. In other words, the integration 
variables are multiplied with a 
certain complex number when the imaginary time is 
shifted by a period. Therefore, the corresponding 
boundary conditions at the ends of the imaginary 
time segment are not of the form corresponding to 
fermionic, or bosonic, variables taken in the 
Matsubara representation at nonzero temperature.
In fact, one part of sites of the models corresponds 
to the integration variables which are subjected 
to the unusual boundary conditions, while the 
variables on the other sites depend on the imaginary 
time conventionally, i.e., as fermions (or bosons). 
Thus a situation, when an ``automorphic'' boundary 
condition is the same for all sites of a chain spin model,  
is generalized. The results of the functional 
integration are obtained in the form of determinants 
of the matrix operators which are regularized by 
means of the generalized zeta-function approach. 
The partition functions of the models and certain 
correlation functions at nonzero temperature are 
obtained explicitly thus demonstrating correctness 
of the functional integral representations proposed.

\end{abstract}

\newpage

\section{Introduction}

The correlation functions of quantum models that are 
solvable via the Bethe ansatz method \cite{qism} can be 
represented, in the thermodynamic limit, as the Fredholm 
determinants of certain linear integral operators. One of such 
{\it determinant} representations has been obtained in 
\cite{lenard} for an equal-time correlator of one-dimensional 
model of ``impenetrable'' bosons, which are described by the 
quantum non-linear Schr\"odinger equation with infinite 
coupling. This result has been generalized 
to the case of correlators with different time arguments 
\cite{kor1}, and also to the case of the $XX$ spin {\it 1/2}
Heisenberg chain \cite{col}. The determinant representations 
of the correlation functions allow to deduce the non-linear
integrable partial differential equations for the correlators 
\cite{qism}, \cite{aaa}. Various determinant representations 
have been deduced in \cite{du}, \cite{ess}, \cite{koj}, 
\cite{kos}, \cite{iz1}, \cite{iz2} (see also Refs. in 
\cite{qism}), for instance, for the temperature correlators 
of quantum non-linear Schr\"odinger equation and for $XXX$ 
and $XXZ$ spin {\it 1/2} Heisenberg chains. It should be 
noticed that $XX$ and $XY$ Heisenberg models still 
continue to attract attention \cite{bil}, \cite{ter}, \cite{jin},
and the multiple integral representations as well as
the determinant representations for the correlators 
in these models are also actively studied \cite{ml},
\cite{pron}. Multiple integration over a set of 
the Grassmann coherent states is used in \cite{iz2}, 
\cite{pron}. 

In its turn, {\it functional integration} (or {\it path 
integration}) technique can be used to calculate the 
correlation functions in various quantum models \cite{hib},
\cite{fad}, \cite{pop}, \cite{camb}, \cite {schulman},
\cite{kl}. The present paper is based on \cite{mal1}, where 
an approach has been proposed to represent the generating 
functional of correlators of $z$-components of spins,
as well as the partition function, in the Heisenberg $XX$-model
by means of functional integrals defined on the variables 
subjected to, so-called, ``automorphic'' boundary conditions.

Approach \cite{mal1} is based on a technical consideration 
carried out in Ref.\cite{alv1} (see also \cite{alv2}) 
which is concerned with the index theory and supersymmetric 
quantum mechanical systems. Path integration is used in 
\cite{alv1} to evaluate traces of those supersymmetric 
quantum mechanical operators which appear in dealing with
various differential geometric indices. In this respect, the 
Ref.\cite{alv1} follows \cite{fad}, \cite{ber}, for the
usage of holomorphic representation of functional integrals
to propose an example of path integral defined on the 
trajectories subjected to non-conventional boundary conditions 
at the ends of the segment of imaginary time.

It has been shown in \cite{mal1} that the generating 
functional of static correlators of $z$-components of local 
spins in the $XX$ Heisenberg magnet can also be represented by 
means of the (Gaussian) functional integrals which are 
defined on the trajectories depending on the imaginary 
time non-conventionally in the sense of \cite{alv1}. 
In other words, the variables of the functional integration 
are multiplied with a certain complex number when the imaginary 
time is shifted by a period, i.e., the variables behave 
``automorphically'' under such shifts. More precisely, the path 
integrals considered in \cite{mal1} are defined for 
the set of variables a part of which is subjected to the 
``automorphic'' boundary conditions, whereas the other part  
satisfy the standard requirements of the fermion/boson-type. 

The point is that a trace of an operator exponential is
evaluated in \cite{mal1}, and the quadratic operator in the
exponent is defined only on the first $m$ sites of the model 
in question ($m\le M$, $M$ is the total number of sites). 
Remind that the $XX$ model considered in \cite{mal1} can 
equivalently be handled in the representation of free 
fermions. Eventually, after a passage from a multiple 
integral over the Grassmann coherent states to the continual 
(i.e., functional) one, it turns out to be possible to define 
the integration domain so that the integration variables are 
``automorphic'' in the imaginary time on the first $m$ sites, 
while they are (anti-)periodic on the other sites. 

The interest to the functional integrals defined on the 
trajectories with ``automorphic'' dependence on the 
imaginary time can also be traced back to \cite{fed},
\cite{gr}. An essential distinction between the formulations
discussed in \cite{mal1} and in \cite{alv1}, \cite{fed} 
consists in the fact that the ``automorphic'' boundary condition 
for the segment of the imaginary time turns out to be 
``inhomogeneous'' spatially since it is valid only for a part 
of sites of the chain model in question. The dependence of the 
integration variables on the imaginary time in \cite{alv1}, 
\cite{fed} is the same for all sites.

The given paper continues \cite{mal1}, and it 
is concerned with carrying of the approach proposed to  
the $XY$ Heisenberg magnet which is equivalent to quasi-free 
fermions (i.e., its Hamiltonian in the fermionic representation
is diagonalized by the Bogoliubov transformation).
It should be noticed that our path integral representations 
do not imply a straightforward implementation of the 
proposal \cite{alv1}: a special restoration of invariance of 
the Lagrangian of the model in question under shifts of the 
imaginary time by a period is required. The method of 
zeta-regularization is used in what follows to 
handle the determinants obtained. The generating functional, 
as well as the partition function of the model, are 
calculated. Certain correlation functions at nonzero 
temperature are obtained explicitly. Thus it is demonstrated 
that the path integration approach proposed admits a 
considerable simplification and enough transparency for the 
problem in question.

From a physical viewpoint, basic ideas of 
{\it zeta-regularization} ($\ze$-regularization) have been 
formulated in \cite{haw}, \cite{dow}, \cite{dew}, \cite{witt}. 
In mathematical literature, usage of 
$\ze$-re\-gu\-la\-riza\-ti\-on is usually traced to \cite{sin}. 
Zeta-regularization turned out to be rather useful in physics 
to calculate, say, the instanton determinants \cite{cor},
\cite{sch1}, the Casimir energy on manifolds \cite{dol}, 
\cite{kant}, as well as the axial and conformal anomalies 
\cite{sch2}. One should be referred to \cite{dew}, \cite{witt},
\cite{sin}, \cite{sch3} for exposition of $\ze$-regularization.

The paper is organized as follows. Section 2 contains outline 
of the problem and basic notations. The representation for the
generating functionals of correlators of $\si^z_{n}$-operators
(and also for the partition functions) in the form given by a
combination of the fermionic functional integrals with 
``automorphic'' boundary conditions is obtained in Section 3 
($\si^z_{n}$ implies the Pauli matrix $\si^z$ at $n$th site). 
Section 4 contains calculation of the functional integrals, 
i.e., obtaining of the answers in the determinant form. The 
most important formulas of $\ze$-regularization are given in 
Section 5. Moreover, the partition function of the $XY$ (and, so, 
of the $XX$) model is calculated in Section 5 with the use of the 
generalized $\ze$-function in the series form. The generalized 
$\ze$-function in the form of a Mellin transform is defined 
in Section 6, and it is used to obtain the regularized 
answers in the form of determinants of finite-dimensional 
matrices which constitute, in their turn, the total generating 
functional. Differentiation of the integrals 
obtained with respect to a parameter is also considered in 
Section 6, and some concrete correlators are calculated. 
Reductions of the answers for the $XY$ model to those of the
$XX$ model are verified.
Discussion in Section 7 concludes the paper.

\section{Outline of the model and notations}

Let us consider the $XY$ Heisenberg magnet of spin {\it 1/2} 
\cite{lb}, \cite{nie} on a periodic chain with the total 
number of sites $M$ (with even $M$). Let $Q(m)$ to denote an 
operator of number of quasi-particles on the first $m$ sites of 
the chain $(m\le M)$. We shall calculate an average of the operator 
exponential $\exp(\al Q(m))$ over the ground state of the model 
(our notations, though conventional, correspond to \cite{col}, 
\cite{iz1}, \cite{iz2}),
$$ 
G(\al, m)\equiv
\l\Phi_0\mid e^{\al Q(m)}\mid\Phi_0\r=
\frac{\Tr(e^{\al Q(m)}e^{-\be H_{XY}})}
{\Tr(e^{-\be H_{XY}})},\quad \al\in\BC,
\eqno(2.1)
$$
where $H_{XY}$ is the Hamiltonian of the $XY$ model, $\be$ 
is inverse temperature ($\be=1/T$), and $\Tr$ means trace of 
operator. The vacuum average (2.1) plays the role of a 
generating functional of static correlators of third components 
of spins of the magnetic chains \cite{col}, \cite{ess}, 
\cite{iz1}, \cite{iz2}, \cite{ml}.

The Hamiltonian of the $XY$ model has the form:
$$
H=H_0+\gamma H_1-hS^z\,,
\eqno(2.2)
$$
where
$$
H_0=-\frac12\sum\limits_{n=1}^M(\sigma^+_n\sigma^-_{n+1} +
\sigma^-_n\sigma^+_{n+1})\,,   
\eqno(2.3)
$$
$$
H_1=-\frac12\sum\limits_{n=1}^M(\sigma^+_n\sigma^+_{n+1} +
\sigma^-_n\sigma^-_{n+1}), \quad
S^z=\frac12\sum\limits_{n=1}^M\sigma^z_n\,,
\eqno(2.4)
$$
where $S^z$ is the total spin, $h$ is an external magnetic 
field ($h\ge0$, \cite{qism}, \cite{col}, \cite{ml}). The 
algebra of the Pauli spin operators on the sites, $\sigma^\al_n$, 
$n\in\{1,\ldots,M\}$, is defined by the commutation relations: 
$$
\big[\,\sigma^\al_k, \sigma^\be_n\,\big] =
2 i\,\delta_{kn}\in{}^{\al\be\gamma}\sigma^\gamma_n,
\quad \sigma^\pm_n=(1/2)(\sigma^x_n\pm i\sigma^y_n),
$$
where $\in^{\al\be\gamma}$ is the totally antisymmetric
symbol, and the indices $\al,\be,\gamma$ acquire the 
``values'' $x,y,z$. Besides, the periodic boundary conditions 
are imposed: $\sigma^\al_{n+M}= \sigma^\al_n$, $\forall n$. 
Real parameter $\gamma$ characterizes anisotropy: we obtain 
the $XX$ Heisenberg magnet at $\gamma=0$.

Let us use the Jordan--Wigner transformation from the variables 
$\sigma^\al_k$ to the canonical fermionic variables $c_k$, 
$c^\dagger_k$:
$$
c_k = \exp\Big(
i\pi\sum\limits^{k-1}_{n=1}\sigma^-_n\sigma^+_n\Big)
\,\sigma^+_k, \qquad
c^\dagger_k = \sigma^-_k\,\exp\Big(
\!-i\pi\sum\limits^{k-1}_{n=1} \sigma^-_n\sigma^+_n\Big).
\eqno(2.5)
$$
The variables $c_k$, $c^\dagger_k$ are subjected to the 
anti-commutation relations:
$$                                                                      
\{c_k, c_n\} = \{c^\dagger_k, c^\dagger_n\}=0, \quad 
\{c_k, c^\dagger_n\}=\dl_{kn}\,,
$$
where the brackets $\{\,,\,\}$ imply anti-commutation.
As a result of the transformation (2.5), the Hamiltonian 
(2.2)--(2.4) acquires the following form in the fermionic 
representation \cite{mc1}:
$$
H=H^+P^+ + H^-P^-,
\eqno(2.6)
$$
$$
H^\pm = -\frac 12\sum\limits^M_{k=1} \Big[c^\dagger_k c_{k+1} 
+c^\dagger_{k+1} c_k + \gamma(c_{k+1}c_k + 
                   c^\dagger_kc^\dagger_{k+1})\Big]
$$
$$
+ h\sum\limits^M_{k=1} c^\dagger_kc_k - hM/2.
\eqno(2.7)
$$
The Hamiltonians $H^\pm$ (2.7) look similiar each to other
except for the choice of the spatial boundary condition
for each of them: the superscripts $\pm$ are chosen in 
correspondence with the boundary conditions for $c_k$, 
$c^\dagger_k$ in the form:
$$
c_{M+1} = \mp c_1, \quad c^\dagger_{M+1} = 
\mp c^\dagger_1.
\eqno(2.8)
$$
We have got in new variables: $Q(m) = \sum\limits^m_{k=1}
c^\dagger_k c_k$ is the number operator of quasi-particles 
on first $m$ sites, the total number of quasi-particles is 
${\cN} \equiv Q(M)$, and the projectors $P^\pm$ in (2.6) are 
defined conventionally: $P^\pm$ $=$ $(1/2)(1\pm(-1)^{\cN})$ 
\cite{mc1}. Operator ${\cN}$ commutes only with $H_0$ (2.3) and 
$S^z$ (2.4) but not with $H_1$ (2.4). The parity operator
$(-1)^{{\cN}}$ anti-commutes with $c^\dagger_k$ and $c_k$,
and it commutes with $H$.

As a result, we obtain the following representation for 
$G(\al, m)$ (2.1) \cite{iz2}:
$$
\begin{array}{c}
G(\al,m)=(2Z)^{-1}(G^+_FZ^+_F + G^-_FZ^-_F + 
                  G^+_BZ^+_B - G^-_BZ^-_B),\\ [0.4cm]
G^\pm_FZ^\pm_F\equiv \Tr\Bigl( e^{\al Q(m)}e^{-\be H^\pm}\Bigr),\\
[0.4cm]
G^\pm_BZ^\pm_B\equiv \Tr\Bigl( e^{\al Q(m)}(-1)^{\cN}e^{-\be H^\pm}
\Bigr)
\end{array}
\eqno(2.9)
$$
(in what follows we omit the subscript $XY$ in the Hamiltonians).
For the partition function $Z$ we have got:
$$
\begin{array}{c}
Z = (1/2)(Z^+_F + Z^-_F + Z^+_B - Z^-_B),\\ [0.4 cm]
Z^\pm_F = \Tr(e^{-\be H^\pm}), \quad Z^\pm_B = 
           \Tr\Bigl( (-1)^{\cN}e^{-\be H^\pm}\Bigr).
\end{array}
\eqno(2.10)
$$

The following observation is discussed in Ref.\cite{alv1},
which is concerned with the index theory and 
supersymmetric quantum mechanics. Let $a$, $a^\d$ be some 
fermionic canonical operators. Let us consider $\CU(1)$-operator 
$Q_\vt\equiv\exp(i\vt a^\d a)$, which acts on $a, a^\d$ as 
follows:
$$
Q_\vt aQ^\d_\vt=e^{-i\vt}a,\quad
Q_\vt a^\d Q^\d_\vt=e^{i\vt}a^\d.
$$
When calculating the trace $\Tr\,(Q_\vt\exp(-\be H))$ in the path 
integral representation ($H$ is a Hamiltonian), it turns out to 
be rather natural to come to a path integral over the variable 
subjected to the ``automorphic'' boundary condition: 
$$
\xi(\tau)=-e^{i\vt}\xi(\tau+\be),
\eqno(2.11)
$$ 
where $\tau$ is imaginary time, $\tau\in [0, \be]$. Indeed,
Eq. (2.11) reminds the definition of an automorphic function 
(automorphic form, \cite{bai}, \cite{saf}):
$$
g^* f\equiv f(gu)=r(g)f(u)\,,
$$
where $f(u)$ is an appropriate function (form), $g$ is an
elememt of a group of transformations acting on the argument
$u$ (thus generating an action of $g^*$ on $f$), and $r(g)$
denotes a representation of $g^*$. Thus, Eq. (2.11)
implies that the integration variable is transformed 
accordingly to a nontrivial representation of $\CU(1)$ when 
$\tau$ is shifted by the period $\be$. Other physical 
(quantum-statistical, in fact) examples of ``automorphic'' 
boundary conditions at the ends of the segment 
$[0, \be]\ni\tau$ can be found in \cite{fed},
where spin {\it 1/2} and spin {\it 1} chain models are 
studied by the method of functional integration.

It is not difficult to note that operator $\exp(\al Q(m))$ 
behaves analogously:
$$
e^{\al Q(m)} c_ne^{-\al Q(m)}=
\left [
\begin{array} {rl}
e^{-\al}c_n, & 1\le n\le m,\\[0.2cm]
        c_n, & m<n\le M,
\end{array}
\right.
\eqno(2.12)
$$
and it is suggestive to use the idea of \cite{alv1} when 
calculating 
$G^\pm_F Z^\pm_F$, $G^\pm_B Z^\pm_B$ (2.9). The ``automorphic'' 
condition arises for all sites in the models considered in 
\cite{alv1}. The peculiarity due to (2.12) is concerned with 
$m\le M$, and the ``automorphic'' condition is expected to 
appear only for a part of sites.

To conclude the section, let us define the coherent states 
using the fermionic operators $c_n$, $c^\d_n$ which possess 
the Fock vacuum $\mid0\r$:
$$
c_n\mid0\r=\l0 \mid c^\d_n=0,\quad
\forall n \in \{1,\ldots,M\},\quad \l 0\mid 0\r=1.
$$
Namely, we define the states
$$
\begin{array}{c}
\displaystyle{
\mid x(a)\r=\exp
\left\{
\sum^M_{k=1} c^\d_k x_k(a)\right\}\mid 0\r
\equiv
\exp (c^\d x(a))\mid 0\r,}\\[0.5cm]
\displaystyle{
\l x^*(a)\mid  =\l 0\mid
\exp
\left\{
\sum^M_{k=1} x^*_k(a)c_k\right\}
\equiv
\l 0\mid \exp (x^*(a)c),}
\end{array}
\eqno(2.13)
$$
where $a$ is discrete index running from 1 to $N$, 
and the shorthand notations are used:
$\sum\limits^M_{k=1}c^\d_k x_k\equiv c^\d x$, 
$\prod\limits^M_{k=1}dx_k\equiv
dx$, etc. In fact, $N$ independent coherent states are 
defined which are labeled by independent complex-valued 
Grassmann parameters $x^*_k(a)$, $x_k(a)$. The following 
relations hold for the states (2.13):
$$
c_k\mid x(a)\r=x_k(a)\mid x(a)\r,\quad
\l x^*(a)\mid c^\d_k=\l x^*(a)\mid x^*_k(a),
$$
$$
\l x^*(a)\mid x(a)\r=\exp (x^*(a)x(a)).
$$

\section{ The functional integral}

Let us turn to the problem of rewriting 
$G^\pm_F Z^\pm_F$, $G^\pm_B Z^\pm_B$ (2.9) and 
$Z^\pm_F$, $Z^\pm_B$ (2.10) in the form of functional 
integrals. For definitness, let us consider 
$G^\pm_FZ^\pm_F$:
$$
G^\pm_F Z^\pm_F=
\int dz\,dz^* e^{z^*z}
\l z^*| e^{\al Q(m)} e^{-\be H^\pm}|z\r,
\eqno(3.1)
$$
where it is understood that the trace of operator is 
calculated as the integral over the anti-commuting variables 
\cite{ber}, \cite{yr}, and the coherent states $\l z^*|,|z\r$ 
are defined analogously to (2.13):
$$
\l z^*|=\l 0|\exp(z^* c),\quad
|z\r=\exp(c^\d z)| 0\r.
$$
In order to go over to the path integral, let us divide
the segment $[0,\be]$ into $N$ parts of the length $\be/ N$,
and let us represent $\exp(-\be H^\pm)$ as a product of $N$ 
identical exponentials. Inserting $N$ decompositions of unity 
between the exponentials, let us transform (3.1) into
$$
\begin{array}{c}
\displaystyle{
G^\pm_F Z^\pm_F=\int dz\,dz^*
\prod\limits^N_{a=1} dx^*(a) dx(a)\exp
\left(z^*z-\sum^N_{a=1} x^*(a)x(a)\right)}\\[0.5cm]
\displaystyle{
\times
\l z^*| e^{\al Q(m)}|x(1)\r \l x^*(1)|
e^{-\frac\be N H^\pm}|x(2)\r\ld
\l x^*(N)| e^{-\frac\be NH^\pm}|z\r},
\end{array}
\eqno(3.2)
$$
where $\mid x(a)\r$ and $\l x^*(a)\mid$ are defined in (2.13).

Using the properties of the coherent states we evaluate the
following averages:
$$
\l z^*| e^{\al Q(m)}| x(1)\r=
\exp\Biggl(
e^\al \sum\limits^m_{k=1} 
z^*_k x_k(1)+\sum^M_{k=m+1} z^*_k x_k(1)
\Biggr),
\eqno(3.3)
$$
$$
\l x^*(a)| e^{-\frac\be NH^\pm}| x(a+1)\r\,\,
{\stackrel{\displaystyle{\simeq}}{_{N\gg1}}}
$$
$$
\simeq \exp\left( 
x^*(a)x(a+1)-\frac\be N\,H^\pm (x^*, x \mid a)\right),
\eqno(3.4)
$$
where
$$
H^\pm(x^*, x \mid a) \equiv H^\pm_0(x^*, x \mid a)
          +\gamma H^\pm_1(x^*, x \mid a)
$$
$$
  +h\sum^M_{k=1}x^*_k(a)x_k(a+1)-\frac{hM}2,
$$
and
$$
H^\pm_0(x^*, x \mid a) \equiv
-\frac12 
\sum^M_{k=1}(x^*_k(a)x_{k+1}(a+1)+x^*_{k+1}(a)x_k(a+1)),
$$
$$
H^\pm_1(x^*, x \mid a) \equiv
-\frac12 
\sum^M_{k=1}(x_{k+1}(a)x_k(a+1)+x^*_k(a)x^*_{k+1}(a+1)).
$$

Inserting (3.3), (3.4) into (3.2), we obtain:
$$
G^\pm_F Z^\pm_F=\int dz\,dz^*
\prod^N_{a=1} dx^*(a)dx(a)\,\exp
\bigg\{
\sum^m_{k=1} z^*_k(z_k+e^\al x_k(1))
$$
$$
+\sum^M_{k=m+1}z^*_k(z_k+x_k(1))+
x^*(1)(x(2)-x(1))+\ld + x^*(N)(z-x(N))
$$
$$
-\frac\be N \Bigl(H^\pm(x^*, x \mid 1)+\ld+
                 H^\pm (x^*, z \mid N)\Bigr)
\bigg\}.
\eqno(3.5)
$$
Let us introduce the notations $x_k(N+1)$ and $x^*_k(0)$
as follows: $x_k(N+1)\equiv z_k$ ($\forall k$), 
$x_k^*(0)\equiv e^\al z_k^*$ (for $1\le k\le m$) or 
$x_k^*(0)\equiv z_k^*$ (for $m<k\le M$).
Further, we impose the boundary conditions with respect
to the counting parameter $a$:
$$
\begin{array} {cll}
x_k(0)=&-e^{-\al}x_k(N+1), & 1\le k\le m,\\[0.2cm]
x_k(0)=&-x_k(N+1),  & m<k\le M,
\end{array}
$$
and perform the transition $N\to\infty$. As a result, the
discrete index $a$ varying from $1$ to $N$ then becomes a
continuous argument $\tau\in [0, \be]$ (imaginary time;
see \cite{ber}, \cite{pop}). The right-hand side of
(3.5) then becomes the integral:
$$
\int \prod_{\tau\in[0,\be]} dx^*(\tau)dx(\tau)\exp
\left(\int\limits^\be_0 \CL(\tau)d\tau\right),
\eqno(3.6)
$$
where $\CL(\tau)$ denotes the Lagrangian:
$$
{\CL}(\tau)= 
\sum\limits_{k=1}^M x_k^*(\tau)\frac{d x_k(\tau)}{d\tau}-
                  H^\pm (x^*, x \mid \tau),
\eqno(3.7)
$$
and the functional variables $x_k(\tau)$ are 
subjected to the ``automorphic'' (see (2.11)) conditions:
$$
\begin{array} {cll}
x_k(\tau)=&-e^{-\al}x_k(\tau+\be), & 1\le k\le m,
\\[0.2cm]
x_k(\tau)=&-x_k(\tau+\be), & m<k\le M.
\end{array}
\eqno(3.8)
$$
Generally speaking, the fields $x^*_k(\tau)$ are 
independent integration variables. It is convenient to 
subject $x^*_k(\tau)$ to a requirement analogous to (3.8) 
but with $e^\al$ instead of $e^{-\al}$.

The derivation of the representation (3.6)--(3.8) follows 
\cite{alv1} strictly, and it does not take into account 
the peculiar character of our problem: the conditions (3.8) 
characterize two independent sets of sites. Therefore, 
the following circumstance becomes essential, which is new 
in comparison with \cite{alv1}, \cite{fed}.

It can be assumed that certain representations of the group of
shifts of $\tau$ by the period $\be$, i.e., $\tau\to\tau+\be$,
are defined by conventinal (anti-)periodicity rules
$x_k(\tau)=\pm x_k(\tau+\be)$, $k\in \{1, \ld, M\}$, as well 
as by the conditions (3.8). The action functional of the model,
$\int\limits^\be_0\CL(\tau)d\tau$, in the exponent of (3.6) is 
well-defined provided the Lagrangian $\CL(\tau)$ is invariant 
under the shifts of $\tau$. Such invariance takes place for 
conventinal boundary conditions provided $\CL(\tau)$ is even in 
powers of the fields.

Let us use (3.8) to calculate the variation $\dl\CL(\tau)$
at $m < M$:
$$
\dl\CL(\tau)\equiv\CL(\tau+\be)-\CL(\tau)=
$$
$$
\begin{array}{c}
\displaystyle{
=\frac12 \Bigl[
(e^\al-1)\Bigl( x^*_{m+1}(\tau)x_m(\tau)+x^*_M(\tau)
x_{M+1}(\tau)\Bigr)\Bigr.}\\[0.5cm]
\qquad\qquad\qquad\qquad\Bigl.
+(e^{-\al}-1)\Bigl( x^*_m(\tau)x_{m+1}(\tau)+
          x^*_{M+1}(\tau)x_M(\tau)\Bigr)\Bigr]
          \end{array}
$$
$$
+\frac{\gamma}{2}\sum\limits^{m-1}_{k=1}
\Bigl((e^{2\al}-1)x_{k+1}(\tau)x_k(\tau) + 
  (e^{-2\al}-1) x^*_k(\tau)x^*_{k+1}(\tau)\Bigr)
$$
$$
\begin{array}{c}
\displaystyle{
+\frac{\gamma}{2}\Bigl[(e^\al-1)\Bigl(
x_{m+1}(\tau)x_m(\tau)+x_{M+1}(\tau)x_M(\tau)\Bigr)\Bigr.}\\
[0.5cm]
\qquad\qquad\qquad\qquad\Bigl.+(e^{-\al}-1)\Bigl(
x^*_{m}(\tau)x^*_{m+1}(\tau)+x^*_M(\tau)x^*_{M+1}(\tau)
\Bigr)\Bigr].
\end{array}
$$
The origin of $\dl\CL(\tau)$ is clear: the quadratic 
forms $H^{\pm}_0$ are invariant under the replacement
$$
x_k\to \pm e^\al x_k,\quad
x^*_k\to\pm e^{-\al} x^*_k,
\eqno(3.9)
$$
provided (3.9) is carried out at each site 
$k\in\{1,\ld,M\}$ (in this case (3.9) looks like a 
homogeneous ``gauge'' transformation), and they are not 
invariant provided (3.9) is valid only for a subset of 
$\{1,\ld,M\}$ (a nonhomogeneous transformation). However,
the forms $H^\pm_1$ are not invariant even for a
homogeneous transformation (3.9). The condition (3.8) implies 
a nonhomogeneous representation of the shifts 
$\tau\to\tau+\be$, and, thus, the invariance turns out to 
be broken for $\CL(\tau)$ (3.7). However, this symmetry can 
straightforwardly be restored as follows: one should replace 
$H^\pm(x^*, x \mid \tau)$ in the limiting formula (3.6) by 
another form ${\w H}^\pm (\tau)
\equiv{\widetilde H}^\pm(x^*, x \mid \tau)$ of the 
following type (we omit the superscript $^\pm$ at $\w H$):
$$
{\w H}(\tau) = {\w H}_0(\tau)
                      +\gamma {\w H}_1(\tau)
 + h\sum\limits^{M}_{k=1}x^*_k(\tau)x_k(\tau) - hM/2,
\eqno(3.10)
$$
where
$$
{\w H}_0(\tau) \equiv
-\frac12 \sum^{M-1}_{k=1}{ }^{^\prime}
\Bigl(x^*_k(\tau)x_{k+1}(\tau) + x^*_{k+1}(\tau)x_k(\tau)\Bigr) 
$$
$$
\begin{array}{c}
\displaystyle{
-\frac12\Bigl(x^*_m(\tau)x_{m+1}(\tau)e^{\al\tau/\be} +
x^*_{m+1}(\tau)x_m(\tau)e^{-\al\tau/\be}}\Bigr.\\[0.5cm]
\qquad\qquad\qquad \Bigl.\displaystyle{
+x^*_M(\tau)x_{M+1}(\tau)e^{-\al\tau/\be} +
x^*_{M+1}(\tau)x_M(\tau)e^{\al\tau/\be}\Bigr)},
\end{array}
$$
$$
{\w H}_1(\tau)\equiv
 -\frac12\sum\limits^{m-1}_{k=1}
\Bigl(x_{k+1}(\tau)x_k(\tau)e^{-2\al\tau/\be} + 
      x^*_k(\tau)x^*_{k+1}(\tau\bigr) e^{2\al\tau/\be}\Bigr) 
$$
$$
 -\frac12\sum\limits^{M-1}_{k=m+1}
  \Bigl(x_{k+1}(\tau)x_k(\tau) + 
                    x^*_k(\tau)x^*_{k+1}(\tau)\Bigr) 
$$
$$
\begin{array}{c}
\displaystyle{
-\frac12\Bigl(x_{m+1}(\tau)x_m(\tau)e^{-\al\tau/\be} +
x^*_m(\tau)x^*_{m+1}(\tau)e^{\al\tau/\be}}\Bigr.
\\[0.5cm]
\qquad\qquad\qquad+\Bigl.x_{M+1}(\tau)x_M(\tau)e^{-\al\tau/\be} +
x^*_M(\tau)x^*_{M+1}(\tau)e^{\al\tau/\be}\Bigr),
\end{array}
$$
where the prime at $\sum\limits$ implies that $k = m$
is skipped in summation. The Lagrangian ${\w {\CL}}(\tau)$ 
takes the form:
$$
{\w {\CL}}(\tau) \equiv \sum\limits^{M}_{k=1} x^*_k(\tau)
\frac{dx_k(\tau)}{d\tau} - {\w H} (x^*, x \mid \tau).
\eqno(3.11)
$$
Equation (3.11) differs from an expected expression 
since it contains the exponential factors 
$\exp(\pm\al\tau/\be)$. The exponentials mentioned ensure 
periodicity of ${\w {\CL}} (\tau)$ with respect to 
the imaginary time \cite{mal1}. Notice that in \cite{alv1} 
and \cite{fed} the ``automorphicity'' conditions 
are the same at each site, and thus the necessity in the 
``compensating'' factors is absent.

The Lagrangian (3.11) is invariant under $\tau\to\tau+\be$, 
the integration measure in (3.6) is also invariant, and, 
finally, we obtain:
$$
G^\pm_FZ^\pm_F
=\int \prod_{\tau\in[0,\be]} dx^*(\tau) dx(\tau)
\exp \left(\int\limits^\be_0\w{\CL}(\tau)d\tau\right),
\eqno(3.12)
$$
where $\w{\CL}(\tau)$ is is given by (3.10), (3.11).
For $G^\pm_B Z^\pm_B$ the functional integral representation
has the same form (3.12). But because of the presence of the
parity operator
$(-1)^{\cN}$ under the trace symbol in (2.9), we get
the corresponding boundary condition in another,
in comparison with (3.8), form:
$$
\begin{array} {cll}
x_k(\tau)=&e^{-\al}x_k(\tau+\be), & 1\le k\le m,
\\[0.2cm]
x_k(\tau)=&x_k(\tau+\be), & m<k\le M
\end{array}
\eqno(3.13)
$$
(i.e., two minus are reversed). Substitution of $G^\pm_F Z^\pm_F$, 
$G^\pm_B Z^\pm_B$, represented by the integrals of the type of 
(3.12) which are supplied with the appropriate boundary
conditions, into (2.9) gives the desired functional
integral representation for the generating functional (2.1). 

The main statement of the present paper reads that the 
representations for $G^\pm_F Z^\pm_F$ and $G^\pm_B Z^\pm_B$ 
given by (3.12) together with the conditions (3.8) and (3.13),
respectively, are well-defined relations which lead to correct
expressions for the correlation functions. Actual 
calculation below is to argue this assertion. We formally 
consider (3.8) and (3.13) as ``automorphic'' boundary 
conditions to distinguish them from more conventional rules 
known for fermions at $\al=i\,2\pi (k+\frac12)$ for (3.8),
or for bosons at $\al=i\,2\pi k$ for (3.13), $k\in\BZ$.

\section{Calculation of the functional integrals}

In order to proceed with calculation of the integral (3.12), 
let us pass to the momentum representation:
$$
\begin{array}{rll}                                              
\displaystyle{
x_k(\tau)=} &
\displaystyle{
(\be M)^{-1/2}e^{-i\pi/4}\sum\limits_{p}
e^{i(\omega\tau-i\frac{\al}{\be}\tau+qk)}x_p,} & 
 1\le k\le m, \\[0.4cm]
\displaystyle{
x_k(\tau)=} &
\displaystyle{
(\be M)^{-1/2}e^{-i\pi/4}\sum\limits_{p}
e^{i(\omega\tau+qk)}x_p,}  & m<k\le M,
\end{array}
\eqno(4.1)
$$
where summation goes over the formal 2-momenta $p=(\omega,q)$:
$\omega$ implies the Matsubara frequencies $\omega=\pi T(2n+1)$, 
$n\in{\BZ}$ (we continue to consider $G^\pm_F Z^\pm_F$), and 
the quasi-momenta $q$ take their values in the sets $X^\pm$
\cite{iz2},
$$
\begin{array} {rl}
X^+=&\left\{q=-\pi+
\displaystyle{\frac{\pi(2l-1)}{M}}\,\, \Bigg |\,\,l=1,\dots ,M 
\right\},
\\[0.4cm]
X^-=&\left\{q=-\pi+
\displaystyle{\frac{2\pi l}{M}}\,\,\Bigg |\,\,l=1,\dots ,M 
\right\}.
\end{array}
\eqno(4.2)
$$
Two sets (4.2) are in correspondence with two ``spatial''
boundary conditions (2.8). As a result of the condition (3.8),
the Fermi frequencies in the Fourier expansions for the
first $m$ sites are shifted by a purely imaginary number. However, 
the summation index in (4.1) can obviously taken as $\om$. 
We use (4.1) in (3.12) and obtain the following representation:
$$
G^\pm_FZ^\pm_F = 
\int \prod\limits_{p}dx^*_pdx_p\exp S^\pm_F (\al),
\eqno(4.3)
$$
$$
S^\pm_F (\al) = \frac{\al m}{2} + 
             \frac{1}{2}\sum\limits_{p}(x^*_p, x_{-p})
\left (\begin{array}{cc}
i\omega-\varepsilon_q & -\Gamma_q \\
-\Gamma_q & i\omega+\varepsilon_q
\end{array}\right)
\left(\!\begin{array}{c}
x_{p} \\
x^*_{-p}
\end{array}\!\right)
$$
$$
+\frac{\al}{2\be}\sum\limits_{p_1,p_2}(x^*_{p_1},x_{-p_1})
\left(\begin{array}{cc}
{\w Q}_{p_1p_2} & 0 \\
0 & -{\w Q}_{p_1p_2}
\end{array}\right)
\left(\!\begin{array}{c}
x_{p_2} \\
x^*_{-p_2}
\end{array}\!\right)\,,
\eqno(4.4)
$$
where the notations are defined:
$$
\begin{array}{c}
\displaystyle{
\varepsilon_q = h - \cos q, 
\quad \Gamma_q = \gamma\sin q, \quad
\w Q_{p_1p_2} = 
\delta_{\omega_1\omega_2}Q_{q_1q_2}},\\
[0.5cm]
\displaystyle{
Q_{pq} = \frac{1}{M}
\frac{\sin\frac{m}{2}(p-q)}{\sin\frac{p-q}2}}\,.
\end{array}
\eqno(4.5)
$$

All the quasi-momenta take independently their 
values in $X^+$ or $X^-$. It has to be pointed out that 
expression $S^\pm_F(\al)$ (4.4) should be viewed a bit formally 
in the case when quasi-momenta belong to the set $X^-$. 
Indeed, when $q\in X^-$, the corresponding Fourier coefficients
$x_{\omega, q}$ and $x^*_{\omega, q}$ require a separate 
consideration at $q=0$ or $q=\pi$. All the necessary explanations 
can be found in \cite{mc1} (see also \cite{iz2}). We shall 
assume that when $q=0$ or $q=\pi$, the corresponding value 
of the argument $-q$ has to be taken equal also to zero or 
to $\pi$, respectively. Extra term $\al m/2$ arises in the matrix 
representation $S^\pm_F(\al)$ (4.4) as a result of an additional
assumption about a {\it fermionic} character of the 
anti-commutation relations for the coefficients $x^*_{p_1}$, 
$x_{p_2}$. Compulsory character of such assumption is due to 
the fact that the presence of this term plays an important 
role for obtaining the correct answers for the correlation 
functions. 
Calculation for $G^\pm_BZ^\pm_B$ (see (2.9)) is carried out 
just analogously to that above, but the presence of $(-1)^{\cN}$ 
under the trace symbol results in the boundary conditions (3.13).
Eventually, summation in the corresponding quadratic 
form $S^\pm_B (\al)$ goes over the bosonic frequencies 
$\omega=2\pi Tn, n\in {\BZ}$, and $G^\pm_B Z^\pm_B$ looks 
similarly to (4.3), (4.4) provided all the frequencies 
are replaced.

Using (3.10)--(3.12) at $\gamma=0$, we obtain the
corresponding expression in the $XX$-case (Eqs.(4.1) are
taken without the factors $e^{-i \pi/4}$):
$$
\begin{array}{c}
\displaystyle{
G^\pm_FZ^\pm_F=\int\prod_{p} dx^*_p dx_p \exp
(S_F^\pm(\al))},\\[0.5cm]
\displaystyle{
S_F^\pm(\al)=\sum_p(i\om-\ep_q)x^*_px_p+\frac\al\be
\sum_{p_1, p_2}
         {\w Q}_{p_1 p_2} 
                x^*_{p_1}x_{p_2}+\frac{\be Mh}2},
\end{array}
\eqno(4.6)
$$
where $\ep_q=h-\cos q$ is the band energy of the 
quasi-particles defined above. When $m=M$, the matrix
$Q_{q_1 q_2}$ (4.5) becomes a Kronecker symbol, and 
therefore $S_F^\pm(\al)$ (4.6) is transformed into a 
free fermionic action with shifted chemical potential.

Let us consider $G^+_FZ^+_F$, given by (4.3), (4.4),
and let us carry out the Bogoliubov transformation 
in $S^+_F(\al)$ with $2\times 2$ matrix ${\bf g}_\theta$
\cite{mc1}, \cite{iz2}:
$$
(x^*_p, x_{-p})=(y^*_p, y_{-p})\,{\bf g}_\theta,
\eqno(4.7)
$$
where
$$
{\bf g}_\theta \equiv 
   \exp\bigl(-i\,\frac{\theta_q}{2}\sigma^2 \bigr)
\in SU(2), \quad \tan\theta_q = 
-\frac{\Gamma_q}{\varepsilon_q}, \quad 
\theta_{-q}=-\theta_q.
$$
Here it is assumed that $\sigma^\al$ ($\al$ takes 
values 1, 2, 3) without a lower index implies 
just a Pauli matrix but not a spin operator on a site. 
Eventually, we obtain the action
$$
\w S^\pm_F(\al) = \frac{\al m}{2} + 
       \frac12\sum\limits_{p}(y^*_p, y_{-p}) \Bigl[ 
         i\omega - E_q\sigma^3\Bigr]
  \left(\!\begin{array}{c} y_p \\
                      y^*_{-p}\end{array}\!\right) 
$$
$$
 +\frac{\al}{2\be}\sum\limits_{p_1, p_2}(y^*_{p_1}, y_{-p_1})
\,{\bf g}_{\theta_1}\Bigl(\w Q_{p_1p_2}\sigma^3\Bigr) 
          \,{\bf g}^{-1}_{\theta_2}
\left(\!\begin{array}{c} y_{p_2} \\
                        y^*_{-p_2}
\end{array}\!\right),
\eqno(4.8)
$$
where $\theta_i \equiv\theta_{q_i} \quad (i=1,2)$, and 
$E_q \equiv (\varepsilon^2_q + \Gamma^2_q)^{1/2}$. 
When $q\in X^-$, the Bogoliubov transformation is carried 
out by means of (4.7) at $q\ne 0, \pi$, while at $q=0, \pi$ 
the remark made after Eq.(4.5) has to be taken into account. 
The point is that, accordingly to \cite{mc1},
we put at $q=0, \pi$:
$$
y_{\om, 0}=x_{\om, 0}, \quad y^*_{\om, 0}=x^*_{\om, 0};
\quad y_{\om, \pi}=x_{\om, \pi}, \quad 
y^*_{\om, \pi}=x^*_{\om, \pi},
$$
and
$$
E_0=h-1=\varepsilon_0, \qquad E_\pi=h+1=\varepsilon_\pi.
$$
In both cases, $q\in X^+$ and $q\in X^-$, the final answer 
is written as a single relation (4.8), where it is assumed 
that $q$ and $-q$ are understood appropriately. 

Substitution of (4.8) into (4.3) (with the change of the 
integration measure) leads \cite{ber} to the answer:
$$
\begin{array}{c}
G^\pm_FZ^\pm_F = e^{\al m/2}\Det^{1/2}A(\al), \\ [0.5cm]
A_{p_1p_2}(\al) \equiv 
\Bigl[ i\omega_F - E_{q_1}\sigma^3\Bigr] \delta_{p_1 p_2} +
\displaystyle{
\frac{\al}{\be}}\,{\bf g}_{\theta_1}
\Bigl(\w Q_{p_1p_2}\sigma^3\Bigr)
\,{\bf g}^{-1}_{\theta_2}.
\end{array}
\eqno(4.9)
$$
Calculation for $G^\pm_BZ^\pm_B$ (2.9) is carried out just 
analogously to that above, and the final answer for 
$G^\pm_B Z^\pm_B$ looks like (4.9) provided all the 
frequencies are replaced:
$$
G^\pm_B Z^\pm_B = e^{\al m/2}\Det^{1/2} 
\left\{ 
\Bigl[ i\omega_B - E_{q_1}\sigma^3\Bigr] \delta_{p_1 p_2} +
\frac{\al}{\be}\,{\bf g}_{\theta_1}
\Bigl(\w Q_{p_1p_2}\sigma^3\Bigr)
\,{\bf g}^{-1}_{\theta_2}
\right\}.
\eqno(4.10)
$$
The partition function of the $XY$-model is given by (2.10) where
$Z^\pm_F$, $Z^\pm_B$ are given as:
$$
\begin{array}{l}
\displaystyle{
Z^\pm_F\equiv \Tr(e^{-\be H^\pm_{XY}})
              =e^{-\be E^\pm_0}
   \Det\Bigl[(i\om_F-E_q)\dl_{pp'}\Bigr]},\\[0.3cm]
\displaystyle{
Z^\pm_B\equiv \Tr\Bigl((-1)^{\cN}e^{-\be H^\pm_{XY}}\Bigr)=
               e^{-\be E^\pm_0}
   \Det\Bigl[(i\om_B-E_q)\dl_{pp'}\Bigr]},
\end{array}
\eqno(4.11)
$$
where (see \cite{iz2})
$$
E^\pm_0\equiv -\frac12 \sum_{q\in X^\pm} E_q,\quad
E_q=(\ep^2_q+\ga^2\sin^2 q)^{1/2}.
$$

Let use the integral representation (4.6)  
to obtain the following formal answers in the $XX$-case \cite{ber}:
$$
\begin{array}{c}
\displaystyle{
G^\pm_FZ^\pm_F=e^{\be Mh/2}\Det
\Bigl[(-i\om_F+\ep_q)\dl_{pp'}-\frac\al\be\,
{\w Q}_{p p'}\Bigr] },\\ [0.5cm]
\displaystyle{
G^\pm_BZ^\pm_B=e^{\be Mh/2}\Det
\Bigl[(-i\om_B+\ep_q)\dl_{pp'}-\frac\al\be\,
{\w Q}_{p p'}\Bigr]},
\end{array}
\eqno(4.12)
$$
and
$$
\begin{array}{l}
\displaystyle{
Z^\pm_F=e^{\be Mh/2}\Det\bigl[(i\om_F-\ep_q)\dl_{pp'}\bigr]},
\\ [0.5cm]
Z^\pm_B=e^{\be Mh/2}\Det\bigl[(i\om_B-\ep_q)\dl_{pp'}\bigr].
\end{array}
\eqno(4.13)
$$
In Eqs.(4.9)--(4.13) we use the notations $\om_F$ and 
$\om_B$ to stress the character, e.g., fermionic or bosonic, of 
the corresponding frequencies. The symbol `Det' denotes 
determinants of infinite-dimensional matrices, while $`\det$' 
is reserved for conventional matrices. Derivations of
the representations (4.12) for $G^\pm_FZ^\pm_F$ and 
$G^\pm_BZ^\pm_B$, as well as of (4.13) for $Z^\pm_F$ and
$Z^\pm_B$, are carried out analogously. It is convenient 
to denote the 
matrix operators, which appear in (4.9), (4.10), (4.12), as 
$A(\al)\equiv A_\al$, while those in (4.11), (4.13) -- as $A$.

\section{Zeta-regularization 
          ({\large Zeta-functions in the series form})}

We shall use $\z$-regularization \cite{sch3} in order to 
assign meaning to the determinants in (4.9)--(4.13). We shall 
begin with the introductory notes. Usually, a generalized 
$\ze$-function is related to an elliptic operator. Precisely, 
let $\CA$ be a non-negative elliptic operator of order $p>0$ 
on a compact $d$-dimensional smooth manifold. Let its 
eigen-values $\la_n$ being ennumerated by the multi-index 
$n$. The series
$$
\ze(s\mid\CA)=\sum_{\la_n\ne0}(\la_n)^{-s},
\eqno(5.1)
$$
which is convergent at $\RE s>d/p$, defines the generalized
$\z$-function of the operator $\CA$, $\ze(s\mid\CA)$. 
This series defines $\ze(s\mid\CA)$ as the meromorphic 
function of the variable $s\in\BC$, which can be analytically 
continued to $s=0$. The formal relation
$$
\lim_{s\to0}\frac{d\ze}{ds}(s\mid\CA)=\lim_{s\to0}
\Biggl[-\sum_{\la_n\ne0}\frac{\log\la_n}{(\la_n)^s}\Biggr]=-\log
\Biggl(\prod_{\la_n\ne0}\la_n\Biggr)
$$
allows to define a regularized determinant of $\CA$ as follows:
$$
\log\Det \CA=-\lim_{s\to0}\frac d{ds}\,\ze(s\mid \CA)
\eqno(
5.2)
$$
The Riemann $\ze$-function,
$$
\ze(s)=\sum^\infty_{n=1}n^{-s},
\quad \RE s>1,
\eqno(5.3)
$$
and the generalized $\ze$-function,
$$
\ze(s,\al)=\sum^\infty_{n=0}(n+\al)^{-s},\quad
\al\ne0,-1,-2,\ld,
\eqno(5.4)
$$
are meromorphic in $s$, have a simple pole at $s=1$ 
with residue 1, and possess a continuation at $s=0$ 
\cite{mag} (see also \cite{tit}). These functions can formally 
be considered as particular cases of the series (5.1). Notice that 
$\ze\left(s,\frac12\right)$ is the Gurvitz $\ze$-function 
\cite{mag}, and $\ze(s,1)=\ze(s)$.

Starting with (5.1), one can represent $\ze(s\mid\CA)$ as a Mellin
transform:
$$
\ze(s\mid\CA)= \frac1{\Ga(s)}\int\limits^\infty_{0}t^{s-1}
[\Tr(e^{-\CA t})-\dim(\ker\CA)]dt.
\eqno(5.5)
$$
The integral (5.5) is defined at sufficiently large positive
$\RE s$ (precisely, at $\RE s>d/p)$; for other $\RE s$
its analytic continuation is required. Equation (5.5)
can be related \cite{sch2}, \cite{sch3} to the definition of 
$\Det\CA$ by means of the proper time regularization \cite{fok},
\cite{jul}:
$$
\log\frac{\Det\CA}{\Det\CA_0}=\Tr
\Biggl[\int\limits^\infty_0(e^{-\CA_0t}-
                        e^{-\CA t})\frac{dt}t\Biggr].
\eqno(5.6)
$$
The definitions of $\log\Det{\cal A}$ by means of (5.2), 
(5.5), and by means of (5.6)
coincide up to an infinite additive constant.

Now one can pass to calculation of the determinants (4.11). 
Let us define the following series, which can be expresed 
through $\ze(s,\al)$ (5.4):
$$
\begin{array}{c}
\displaystyle{
\ze^\pm_F(s\mid A)\equiv \sum_{\om_F,q\in
X^\pm}(i\om_F-E_q)^{-s}}\\[0.5cm]
\displaystyle{
=\left(\frac{\be}{2\pi i}\right)^s
\sum_{q\in X^\pm}
\left[
\ze\Biggl(
s,\frac12+i\frac{\be E_q}{2\pi}\Biggr)+
(-1)^s\ze
\Biggl(s,\frac12-i\frac{\be E_q}{2\pi}\Biggr)\right]},
\end{array}
\eqno(5.7)
$$
$$
\begin{array}{c}
\displaystyle{
\ze^\pm_B(s\mid A)\equiv \sum_{\om_B,q\in
X^\pm}(i\om_B-E_q)^{-s}}\\[0.5cm]
\displaystyle{
=\left(\frac{\be}{2\pi i}\right)^s \sum_{q\in X^\pm}
\left[
\ze\Biggl(
s,i\frac{\be E_q}{2\pi}\Biggr)+
(-1)^s\ze
\Biggl(s,-i\frac{\be E_q}{2\pi}\Biggr)\right]-\sum_{q\in
X^\pm}(-E_q)^{-s}}.
\end{array}
\eqno(5.8)
$$
The series $\ze^\pm_F(s\mid A)$ and $\ze^\pm_B(s\mid A)$ 
should be considered as the generalized $\ze$-functions of 
the diagonal operators $A$ (see (4.11), (4.13)) in the 
series form (5.1).
The analytic continuations for $\ze(s,z)$ are known \cite{mag}:
$$
\ze(0,z)=\frac12-z,\quad
\ze'(0,z)=\log\frac{\Ga(z)}{(2\pi)^{1/2}},
\eqno(5.9)
$$
and they lead to the following answers:
$$
\begin{array}{c}
\displaystyle{
-\lim_{s\to0}\frac d{ds}\,\ze^\pm_F(s\mid A)=\sum_{q\in
X^\pm}\log (1+e^{c\be E_q})},\\[0.5cm]
\displaystyle{
-\lim_{s\to0}\frac d{ds}\,\ze^\pm_B(s\mid A)=\sum_{q\in
X^\pm}\log (1-e^{c\be E_q})},
\end{array}
\eqno(5.10)
$$
where $c=\pm1$ due to an arbitrariness when differentiating
$(-1)^s$ $=$ $\exp(\pm i\pi s)$. 

Choosing $c=-1$, and combining (5.10) with (5.2), one 
obtains the following relations of the $XY$-model 
\cite{iz1}, \cite{iz2}:
$$
\begin{array}{c}
\displaystyle{
Z^\pm_F=e^{-\be E^\pm_0}\prod_{q\in X^\pm}(1+e^{-\be E_q})
=\prod_{q\in X^\pm}2\cosh\frac{\be E_q}2},\\[0.5cm]
\displaystyle{
Z^\pm_B=e^{-\be E^\pm_0}\prod_{q\in X^\pm}(1-e^{-\be E_q})
=\prod_{q\in X^\pm}2\sinh\frac{\be E_q}2}.
\end{array}
\eqno(5.11)
$$
The total partition function should be calculated accordingly
to (2.10), the free energy is $F=-(1/\be M)\log Z$, while 
arbitrariness in the choice of $c$ does not influence the 
magnetization $M_z=-\cd F/\cd h$ and the entropy 
$S=-\cd F/\cd T$. Specifically, one gets in the thermodynamic 
limit \cite{iz1}:
$$
F=-\frac1{2\pi\be}\int\limits^\pi_0\log (2(1+\cosh\be E_q))dq.
\eqno(5.12)
$$
All the formulas obtained can be reduced at $\ga\to 0$ to those
of the $XX$-model.

\section{Determinants of the operators $A(\al)$}
\subsection{The regularization (Zeta-functions in the integral
form)}

Thus, in the previous section we have defined 
$\ze$-functions of the diagonal operators $A$ given by (4.11), 
(4.13) in the series form. Let us now use (5.5) to calculate 
the regularized determinants of the non-diagonal operators 
$A_\al$ given by (4.9), (4.10), (4.12). For instance, let us 
now proceed with the calculation of $G^\pm_F$ (4.12).

Let us begin with the formal integral
$$
\frac1{\Ga(s)}\int\limits^\infty_0 t^{s-1}\Tr\left[
e^{(i\om_F-\h\ep+\frac\al\be\h Q)t}\right]dt,
\eqno(6.1)
$$
where $\h\ep$ and $\h Q$ imply the matrices in the momentum
space, $\diag\{\ep_q\}$ and $Q_{pq}$ (4.5), accordingly,
while the indices $p$, $q$ run independently over $X^+$ or 
$X^-$. Trace `${\Tr}$' in (6.1) is considered as a matrix
one over the corresponding 2-momenta $p=(\om, q)$ which
label entries of our matrix operators. Convergence of the 
integral (6.1) at the upper bound is respected at 
sufficiently large $h>h_c=1$ ($h_c$ is the critical magnetic 
field \cite{col}). Regularization of the integral is necessary 
at the lower bound.

Let us use the asymptotical relation
$$
\Tr\left[ e^{(i\om_F-\h \ep+\frac\al\be\h
Q)t}\right]\stackrel{\longrightarrow}{_{t \to\,0}}\phi_0,
$$
where $\phi_0$ is an infinite constant 
equal to $\Tr(\dl_{pp'})\equiv\displaystyle{\sum\limits_{\om_F}}
\tr \h\dl$ ($\h\dl$ is a unit $M\times M$ matix). Let us 
define the function $\rho(t)$:
$$
\rho(t)\equiv\Tr\left[ e^{(i\om_F-\h \ep+\frac\al\be\h
Q)t}\right]-\phi_0,\quad
0\le t <1 \,,
\eqno(6.2)
$$
and divide the integral (6.1) into two parts. We 
rewrite (6.1) using (6.2) as follows:
$$
\frac1{\Ga(s)}\int\limits^\infty_1
t^{s-1}\Tr
\left[ e^{(i\om_F-\h \ep+\frac\al\be\h
Q)t}\right] dt+
\frac1{\Ga(s)} \int\limits^1_0 t^{s-1}\rho(t)dt+
                         \frac{\phi_0}{s\Ga(s)}.
\eqno(6.3)
$$
The function $\rho(t)$ is a formal series in powers of $t^n$, 
$n\ge1$. Besides,
$$
\frac1{s\Ga(s)}\simeq 1+\ga s+o(s),\quad \ga=-\psi(1),
\eqno(6.4)
$$
where $\psi(z)=(d/dz)\log\Ga(z)$. Therefore, (6.3), which 
is regular at $s\to0$, defines an analytic continuation of 
(6.1) at any $\RE s\ge0$. It just can be considered as the 
definition of $\ze^\pm_F(s\mid A_\al)$ in the right 
half-plane of $\BC\ni s$.

Let us now consider the constant $\phi_0$ and the 
coefficients which define $\rho(t)$. In our situation all 
these coefficients are given by divergent series, but finite 
(i.e., regularized) values can be assigned to them by means 
of reductions of the series to zeta-functions (5.3), (5.4), 
i.e., to their particular values at special arguments 
\cite{mag}.

First of all, using $\ze(0)=-\frac12$ one obtains:
$$
\phi_0=M\sum_{\BZ}1=M(2\ze(0)+1)=0,
$$
where we can equivalently replace ${\displaystyle 
\sum_{\BZ}}$ and $2\ze(0)+1$ by ${\displaystyle 
\sum_{\BZ+\frac12}}$ and $2\ze\left(0,\frac12\right)$,
respectively
(``$\ze$-regularized measure'' of the set $\BZ$ is zero). 
Further, the divergence of the coefficients at the powers 
of $t$ in $\rho(t)$ is given by the divergent sums
$
\sum\limits_{n\in\BZ+\frac12}n^m.
$
It is reasonable to consider such sums as zeros at 
$m=2k+1$, $k\in \BZ^+$, since $n^m\equiv 
\left(l+\frac12\right)^m$ are odd. If $m=2k$, 
$k\in\BZ^+$, then
$$
\sum_{n\in\BZ+\frac12} n^m=2\sum^\infty_{l=0}
\left(l+\frac12\right)^m= 2\ze
\left(-2k,\frac12\right)=2\left(\frac1{2^{2k}}-1\right)\ze(-2k).
$$
But $\ze(-2k)=0$ at $k\ge1$. It can be concluded that
all ``$\ze$-regularized'' coefficients are zero for 
$\rho(t)$, corresponding to our $A_\al$, 
and, thus, only the first term is relevant in (6.3).

Let use (6.4) to pass from (6.3) to the relation:
$$
-\lim_{s\to0}\frac d{ds}\ze^\pm_F (s\mid A_\al)
=
-\int\limits^\infty_1\tr
\left[ e^{(-\h\ep+\frac\al\be\h Q)t}\right]
\left(\sum_{\om_F}e^{i\om_Ft}\right)\frac{dt}t-
\int\limits^1_0 \rho(t)\frac{dt}t-\ga\phi_0,
\eqno(6.5)
$$
where the Poisson summation formula enables to sum up over
$\om_F$. Then, R.H.S. of (6.5) takes the form:
$$
-\tr\sum^\infty_{k=1}\frac{(-1)^k}k
\left( e^{-\be\h\ep+\al\h Q}\right)^k-\int\limits^1_0
\rho(t)\frac{dt}t-\ga\phi_0
$$
$$
=
\log\det\left({\h\dl}+
 e^{-\be\h\ep+\al\h Q}\right)
     -\int\limits^1_0\rho(t)\frac{dt}t-\ga\phi_0,
$$
where ${\h\dl}$ is a unit $M\times M$ matrix.
Let us also take into account that $\h Q^2=\h Q$, and so,
$e^{\al \h Q}-{\h\dl}=(e^\al-1)\h Q$. Therefore,
$$
G^\pm_F=\frac{\Det\left[
(i\om_F-\ep_q)\dl_{pp'}+\displaystyle{
\frac\al\be}\,{\w Q}_{p p'}\right]}
{\Det[(i\om_F-\ep_q)\dl_{pp'}]}
=
\det\left[\h\dl+(e^\al-1)\h Q({\h\dl}+e^{\be\h\ep})\1\right].
\eqno(6.6)
$$
Additional renormalization of $\rho(t)$ and $\phi_0$ (to zero, 
in fact) is irrelevant for $G^\pm_F$ written as the ratio of the
determinants. However, when $m=M$, the corresponding operator
$A_\al$ becomes diagonal since $\h Q$ becomes a unit matrix
$\h\dl$. In this case, we can use $\ze$-function in the
series form (5.1). Transparent adjusting of (5.7)--(5.10) 
for this case (i.e., for $m=M$) gives the same answer as that
given by (6.5) where $\rho(t)$ and $\phi_0$ are taken zero. 
Further, we proceed similarly to obtain:
$$
G^\pm_B=\det\left[{\h\dl}+{\h M}_B(\al) \right],
\eqno(6.7a)
$$
where we introduced the matrix notation
$$
{\h M}_{F, B} (\al)\equiv 
\frac{(\displaystyle{e^\al}-1) {\h Q} }{{\h\dl}\pm \exp(\be\h\ep)}.
\eqno(6.7b)
$$
The matrices ${\h M}_{F, B} (\al)$ are of the size
$M\times M$; we take $+$ or $-$ in the R.H.S. of (6.7b)
provided the subscript in the L.H.S. is $F$ or $B$,
respectively.

In deriving (6.6), (6.7), we have restricted ourselves to the 
case where $h>1$. For $0< h<1$, the energy $\ep_q$ is not 
strictly positive, and a potential problem of convergency of 
the integral (6.1) at the upper bound arises. It can be shown 
that in this case it is necessary to use the integral 
representation (6.1) to calculate a regularized 
value of $\Det^{1/2}(A_\al A^*_\al)$, where $A^*_\al$
is a complex conjugated to $A_\al$. Consideration
of the corresponding complex ``phase'' of the
determinant of $A_\al$ leads to the same final answer
as that at $h>1$.

Let us illustrate the last statements by the following 
calculation. We are gowing to calculate $\Det(i\om_F+\la)$.
Firts of all, let us use (5.5) to calculate 
$$
\Det^{1/2}(\om_F^2+\la^2)=
{\sqrt{\phantom {.}\!{\rm D}}}{\rm et} 
                (i \om_F +\la)\,\Det(-i\om_F+\la),
$$
where $\la$ is nonzero real number. Let us start with
the representation
$$
\frac1{\Ga(s)}\int\limits^\infty_0 t^{s-1}\Tr\left[
e^{-(\om_F^2+\la^2)t}\right]dt.
$$
We use the estimation:
$$
\Tr\left[\,e^{-(\om_F^2+\la^2)t}\,\right]\,\,
{\stackrel{\displaystyle{\sim}}{_{t\to 0}}}\,\,
\frac{\be}{2(\pi t)^{1/2}},
$$
where the Poisson summation formula is used to re-express
the series 
$$
\sum\limits_{\om_F} \exp\bigl(-\om_F^2 t\bigr)
=\frac{\be}{2 {\sqrt{\pi t}}}
\sum\limits_{\BZ} (-1)^k\exp\left(-\frac{\be^2 k^2}{4 t}\right),
\quad 0\le t<1,
$$
and we define the function $\rho (t)$ (compare with
(6.2)) as follows:
$$
\rho(t)\equiv
\Tr\left[\,e^{-(\om_F^2+\lambda^2)t}\,\right]
-\frac{\be}{2(\pi t)^{1/2}}.
$$
Now, the corresponding generalized $\z$-function of the
diagonal matrix operator 
${\it diag}\,\bigl\{\om_F^2+\la^2 \bigr\}$ takes the form:
$$
\begin{array}{r}
\displaystyle{
\z\Bigl(s\,{\Big |}\,\diag \bigl\{\om_F^2+\la^2 \bigr\}\Bigr)
       \equiv\frac1{\Ga(s)}
\left[ \int\limits^\infty_1 t^{s-1}\Tr
\Bigl(e^{-(\om_F^2+\la^2)t}\Bigr) dt \right.}\\ [0.5cm]
\displaystyle{
\left.+\int\limits^1_0 t^{s-1}\rho(t)dt+
                         \frac{\be}{2{\pi}^{1/2}}\,(s-1/2)^{-1}
\right]}.
\end{array}
\eqno(6.8)
$$
Equation (6.8) defines an analytic continuation of 
$\ze$-function in the right half-plane of ${\BC}\ni s$.

Using (6.8) we proceed with the calculation of 
the logarithm of
the determinant
$\Det^{1/2}(\om_F^2+\la^2)$:
$$
-\frac 12\,\lim_{s\to0}\frac d{ds}\,\ze \Bigl(s\,{\Big |}\,\diag 
\bigl\{\om_F^2+\la^2 \bigr\}\Bigr)= -\frac12 \int\limits^\infty_1
\left(\sum_{\om_F}e^{-\om_F^2 t}\right)
e^{-\lambda^2 t}\,\frac{dt}t-\frac12
\int\limits^1_0 \rho(t)\frac{dt}t +\frac\be{2{\pi}^{1/2}}
$$
$$
= \log(1+e^{-\be|\la|})-\frac\be{4{\pi}^{1/2}}
\left( \int\limits^\infty_1 e^{-\la^2 t}\frac{dt}{t^{3/2}}
+ \int\limits^1_0 \bigl(e^{-\la^2 t}-1\bigr)\frac{dt}{t^{3/2}}
\right)+ \frac\be{2{\pi}^{1/2}}
$$
$$
= \log\Bigl(2\,\ch\frac{\be|\la|}2\Bigr).
\eqno(6.9)
$$
In right-hand side of (6.9) we can relace $|\la|$ by $\la$. 
Thus we have obtained the regularized value for 
the logarithm of
$\Det^{1/2}(\om_F^2+\la^2)$. We determine the ``phase'' 
of $\Det(i\om_F+\la)$ as follows:
$$
\sum\limits_{\om_F}\at\frac{\om_F}\la=
\frac{-i}2\ \sum\limits_{\om_F}\log\frac{\la+i\om_F}{\la-i\om_F}
=-i c\,\frac{\be\la}2, \qquad c=\pm 1,
\eqno(6.10)
$$
where we take into account the fact that 
$\displaystyle{\sum\limits_{\BZ}}1 =2\ze(0)+1=0$, and we use
$$
-\lim_{s\to0}\frac d{ds}\left(\sum\limits_{\om_F}
(i\om_F+\ell)^{-s}\right)=\log(1+e^{-c\be\ell})
$$
at $\ell=\pm\la$. Combining (6.9) and (6.10), we obtain, 
in agreement with the previous calculations (5.7)--(5.10), 
the following answer:
$$
\Det(i\om_F+\la) = 1+e^{c \be \la}\,.
\eqno(6.11)
$$
Therefore, for any sign of the parameter (``energy'') $\la$, 
we can use a freedom in the choice of $c$ so that the 
final result, say, for the eigen-value matrix 
$\widehat\varepsilon =\diag \left\{\varepsilon_q\right\}$ 
with $\varepsilon_q$ (4.5) will be the same at $0< h<1$ as
that for the magnetic field $h>1$. Since our matrix operators 
are diagonalizables, the calculation of (6.11) by means of 
(6.8)--(6.10) justifies our manipulations leading
to (6.6), (6.7) at any $h>0$. For the bosonic 
frequencies $\om_B$ the situation is similiar.

Now, let us turn to the $XY$-case. We shall define the 
regularized determinant of the matrix operator $A(\al)$ 
(4.9) by means of the generalized zeta-function of this 
operator $\zeta (s | A(\al))$ as follows:
$$
\log\Det^{1/2}A(\al) = 
-\frac12 \lim_{s\rightarrow 0}\frac{d}{ds} \zeta (s | A(\al)),
$$
where a standard representation of $\zeta (s | A(\al))$ by 
the Mellin integral (5.5) is meant. Referring to the previous 
calculations concerning the $XX$-case, we shall write the 
answers as follows: 
$$
G^\pm_F = e^{\al m/2}\,\frac{
\Det^{1/2} \left[ 
i\omega_F - {\h E}\otimes\sigma^3 +
\displaystyle{\frac{\al}{\be}}\,{\w g}
\Bigl({\h Q}\otimes\sigma^3\Bigr)\,{\w g}^{-1}
\right]}{\Det^{1/2}\Bigl[i\om_F-{\h E}\otimes\sigma^3\Bigr]}
$$
$$
= {\det}^{1/2}_M(e^{\al{\h Q}})\,{\det}^{1/2}_{2M}
\left\{{\w I} + \frac{{\w g}
\bigl[ {\h Q}\otimes\bigl(\exp(-\al\sigma^3\bigr)- I)\bigr]
\widetilde g^{-1}} 
{{\w I} + \exp(-\be{\h E}\otimes\sigma^3)}\right\},
\eqno(6.12)
$$
\vskip0.3cm
$$
Z^\pm_F ={\det}^{1/2}_{2M} \left\{{\w I} + 
\exp(\be{\h E}\otimes\sigma^3)\right\}.
\eqno(6.13)
$$
The determinants of the infinite-dimensional and 
finite-dimensional matrices are denoted in (6.12), (6.13) as 
`Det' and `det', respectively, while the indices $M$ and 
$2M$ imply the size of the matrices. In (6.12), ${\w I}$ 
denotes a unit $2M\times 2M$ matrix, ${\w g}$ denotes a 
block-diagonal $2M\times 2M$matrix with the blocks on its 
principal diagonal given by the matrices ${\bf g}_\theta$ 
defined in (4.7); $\widehat Q$ and $\widehat E$ are $M\times M$ 
matrices with the entries $Q_{q_1q_2}$ and $E_{q_1}\dl_{q_1q_2}$, 
correspondingly, and $I$ is $2\times 2$ unit matrix. Besides, 
the following relations are used:
$$
\exp\Bigl(-\al\,\widetilde g\bigl(
{\h Q}\otimes\sigma^3\bigr)\,{\w g}^{-1} \Bigr) - {\w I} =
\,{\w g} \Bigl[{\h Q} \otimes\bigl(\exp(-\al\sigma^3)-I\bigr)
\Bigr]\,{\w g}^{-1},
$$
and $\det(\exp\al\widehat Q) = \exp(\al m)$ (see notations in 
(4.5)). The notations for tensor products in (6.12), (6.13) are 
in agreement with the matrix notations in (4.8)--(4.10).

Let us verify some reductions of the relations (6.12), (6.13). 
First of all, let us calculate $Z^\pm_F$ (6.13):
$$
Z^\pm_F = \sqrt{{\prod\limits_{q\in X^\pm}}}(1+e^{\be E_q})
(1+e^{-\be E_q}) = 
\prod\limits_{q\in X^\pm} 2\,\cosh \frac{\be E_q}{2}.
\eqno(6.14)
$$
Further, we shall consider (6.12). When $m=M$, the matrix 
$\h Q$ becomes a unit matrix $\h\dl$, and we obtain at 
$\al = i\pi$ (see (5.11), as well as the definitions (2.9) and
(2.10)):
$$
G^\pm_F = \prod\limits_{q\in X^\pm} \tanh \frac{\be E_q}{2}
= Z^\pm_B/Z^\pm_F 
\eqno(6.15)
$$
Let us now put $m=M$, and $\al$ is arbitrary. Then,
we obtain from (6.12):
$$
G^\pm_F = e^{\al M/2} \sqrt{{\prod\limits_{q\in X^\pm}}}{\det}_2
\left[ I - \frac{(1-\ch\al) I
                    +\sh\al\,{\bf g}_{2\theta}\,\sigma^3}
        {I+\exp(-\be E_q\sigma^3)}\right]
$$
$$
= e^{\al M/2} {\sqrt{\phantom {\Pi } 
\!\!\!\!\prod\limits_{q\in X^\pm}}}
\left(\ch^2\frac{\al}{2} - 
\sh\al\cos\theta_q \th\frac{\be E_q}{2} 
+ \sh^2\frac{\al}{2}\,\th^2\frac{\be E_q}{2}\right),
\eqno(6.16)
$$
where ${\bf g}_{2\theta}$ is the matrix of rotation by 
the angle $2\theta_q$, $I$ is a $2\times 2$ unit
matrix. On the other hand, the result of \cite{iz2} reads:
$$
G^\pm_F = {\det}_M \left({\h\dl} + {\h K}_F (\al) \right),
\qquad {\h K}_F (\al)\equiv (e^\al - 1)\,\frac{\widehat Q}{2}\,
\diag\Bigl\{ 1 - 
          e^{i\theta_q}\th\frac{\be E_q}{2}\Bigr\}.
\eqno(6.17)
$$
Direct calculations of $G^\pm_F$ by means of (6.17) taken
at $m=M$, and by means of (6.16), coincide. Finally, let us take 
$\ga=0$ while $\al$ and $m$ both are arbitrary in (6.12). 
Then we obtain the answer (6.6) for the $XX$-case as follows:
$$
G^\pm_F = 
{\det}^{1/2}_{2M}
\left(\!
\begin{array}{cc} 
\exp(\al {\h Q}) & 0 \\
0 & {\h \dl}
\end{array}\!\right)\,
{\det}^{1/2}_{2M}
\left(\begin{array}{cc}
{\h \dl} + 
\displaystyle{
\frac {\exp(-\al {\h Q})-{\h \dl}} {{\h \dl} + 
\exp(-\be{\h\varepsilon})}} & 0 \\
0 & {\h \dl} + 
\displaystyle{
\frac {\exp(\al {\h Q})-{\h \dl}} {{\h \dl} + 
\exp(\be{\h\varepsilon})}}
\end{array}\right)
$$
$$
={\det}_M\biggl({\h\dl} + \frac {\exp(\al\widehat Q)-{\h\dl}} 
{{\h\dl} + \exp(\be{\h\varepsilon})}\biggr),
\eqno(6.18)
$$
where the matrices ${\h Q}$ and ${\h\dl}$ are defined above, 
and $\widehat \varepsilon = \diag\{\varepsilon_q\}$,
$\varepsilon_q\equiv h-\cos q$.

It is clear from (6.14)--(6.16), (6.18) that the 
representations (6.12), (6.13) are in agreement with the 
results (6.6), (6.7) \cite{mal1}, and they 
reproduce the relations of the $XX$ model \cite{col}, \cite{iz1}, 
\cite{iz2}, correctly. Calculation for $G^\pm_BZ^\pm_B$ is 
carried out just analogously to that above. The final answers 
for $G^\pm_B$ and $Z^\pm_B$ are:
$$
G^\pm_B = e^{\al m/2}\,\frac{
\Det^{1/2} \left[ 
i\omega_B - {\h E}\otimes\sigma^3 +
\displaystyle{\frac{\al}{\be}}\,{\w g}
\Bigl({\h Q}\otimes\sigma^3\Bigr)\,{\w g}^{-1}
\right]}{\Det^{1/2}\Bigl[i\om_B-{\h E}\otimes\sigma^3\Bigr]}
$$
$$
= {\det}^{1/2}_M(e^{\al{\h Q}})\,{\det}^{1/2}_{2M}
\left\{{\w I} +  {\w M}_B (\al) \right\},
\eqno(6.19a)
$$
\vskip0.3cm
$$
Z^\pm_F ={\det}^{1/2}_{2M} \left\{{\w I} - 
\exp(\be{\h E}\otimes\sigma^3)\right\},
\eqno(6.19b)
$$
where
$$
{\w M}_{F,\,B} (\al) \equiv \frac{{\w g} \bigl[ {\h Q}\otimes
\bigl(\exp(-\al\sigma^3\bigr)- I)\bigr]\widetilde g^{-1}} 
{{\w I} \pm \exp(-\be{\h E}\otimes\sigma^3)}.
\eqno(6.19c)
$$
In (6.19c) it is also meant that $+$ or $-$ correspond
to $F$ or $B$, accordingly. We see that the answers
(6.12) and (6.19a) differ only with respect to
the signs at $\exp(\pm\be{\h E}\otimes\sigma^3)$.
Analogously, the results in \cite{iz2} for $G^\pm_F$ and 
$G^\pm_B$ also differ only with respect to the signs in 
front of $\exp(\be E_q)$.

It is appropriate to notice that the representation
$$
G^\pm_F = e^{\al m/2}\Det^{1/2} \left\{{\w I} +\frac{\,\al\,}{\be}
\frac{\,\widetilde g\,\Bigl(\widehat Q\otimes\sigma^3\Bigr)\,
\widetilde g^{-1}\,}
      {i\omega_F-\widehat E\otimes\sigma^3}\right\}
$$
$$
= e^{\al m/2}\exp \left\{\frac12\sum\limits^{\infty}_{k=1}
\frac{(-1)^{k-1}}{k}\,\Tr\!\left[\frac{\,\al\,}{\be}\frac
{\,\widetilde g\,\Bigl(\widehat Q\otimes\sigma^3\Bigr)\,
\widetilde g^{-1}\,} {i\omega_F - 
            \widehat E\otimes\sigma^3}\right]^k\right\}
\eqno(6.20)
$$
(which implies regularization of $\Det A(\al)$ more conventional 
for quantum field theory) leads to the same numerical coefficients
at the powers of $\al$, as Eq.(6.12). For the $XX$-case, (6.20) 
acquires the following form:
$$
G^\pm_F=\Det\left\{{\h\dl}+\frac\al\be\,\frac{\h Q}
                     {i\om_F-{\h \ep}_q}\right\}
= \exp\left\{\sum_{k=1}^{\infty}
\frac{(-1)^{k-1}}k\,\Tr \left[\frac\al\be\,\frac{\h Q}
{i\om_F-{\h\ep}_q}\right]^k\right\}
\eqno(6.21)
$$
In (6.20) and (6.21), ${\w I}$ and ${\h \dl}$ imply the 
corresponding unit operators.

\subsection{ Differentiation of the determinants}

Calculation of the correlation functions by means 
of the generating functional $G(\al, m)$ (2.1) is related
with differentiations of it over $\al$ and $m$ \cite{qism}, 
\cite{kor1}, \cite{col}, \cite{ess}, \cite{kos}, 
\cite{iz1}, \cite{iz2}, \cite{ml}. In fact, in order to 
calculate correlators of $z$-components of spins (the 
operator of third component of spin, $\si_{m}^z$, is defined as 
$\si^z$ at $m$th site), we differentiate $G(\al, m)$ over $\al$ at 
$\al=0$ as follows \cite{qism}, \cite{col}:
$$
\langle Q^n(m)\rangle
=\lim_{\al\to0}\frac{d^n}{d\al^n}\,G(\al, m).
\eqno(6.22)
$$
However, with regard at (6.1), it is suffice to do only 
a first differentiation. The other ones occur as 
usual differentiations of matrices \cite{gan}.

The operators in question, $A(\al)$, are linear in $\al$:
$A(\al)\equiv A_1+\al A_2$. Let us calculate the first 
derivative of $\Det A(\al)$ using the formal integral (6.1):
$$
\frac{(d/d\al)\Det A(\al)}{\Det A(\al)}=-\frac d{d\al}
\left(\int^\infty_0\Tr(e^{A(\al)t})\frac{dt}t\right)
\eqno(6.23)
$$
(the regularization at $t\searrow 0$ is irrelevant for the
differentiation over the parameter). In the spirit of the
Ray--Singer--Schwarz lemma \cite{sch4}, we shall use in
(6.23) the following relation:
$$
\frac d{d\al}\Bigl(\Tr(e^{A(\al)t})\Bigr)
=t\frac d{dt}\,\Tr\Bigl(B(\al)e^{A(\al)t}\Bigr),\quad
B(\al)\equiv A_2 A\1(\al).
$$
Then, the integral over $t$ can be calculated, and we
obtains in the $XX$-case:
$$
\frac{d}{d\al}(\log\Det A_\al ) = \Tr B_\al
$$
$$
=\Tr\left(
\frac{\h Q}\be(i\om_F-\h\ep+\frac\al\be\h Q)\1\right)=
\tr\Bigl(\h Q(1+e^{\be\h\ep-\al\h Q})\1\Bigr),
\eqno(6.24)
$$
where the `fermionic' $A_\al$ (4.12) is meant, and the Cauchy 
formula for matrices \cite{gan} is used to sum up over the 
frequencies. In the $XY$-case, for the matrix operator, say, 
$A_\al$ (4.9), the answer reads:
$$
\frac{d}{d\al}\bigl(\log\Det^{1/2} A_\al\bigr)\,
=\,-\frac12\,
\tr\!\left(\frac
{\widetilde g\,\big(\widehat Q\otimes\sigma^3\big)\,
\widetilde g^{-1}}
  {{\w I}+\exp\big(\!-\be\widehat E\otimes\sigma^3 + \al\, 
    \widetilde g\,(\widehat Q\otimes\sigma^3)\,\widetilde 
g^{-1}\big)} \right).
\eqno(6.25)
$$

Knowing (6.24), (6.25), one can carry out all the 
differentiations required. Thus, we obtain,
\vskip0.3cm
\leftline{a) in the $XX$-case:}

$$
\lim_{\al\to0}\frac{(d/d\al)\Det A_\al}{\Det A_\al}=
\tr\Bigl(\h Q({\h\dl}+e^{\be\h\ep})\1\Bigr),
$$
$$
\eqno(6.26)
$$
$$
\lim_{\al\to0}\frac{(d^2/d\al^2)\Det A_\al}{\Det A_\al}=
\tr\Bigl(\h Q({\h\dl}+e^{\be\h\ep})\1\Bigr)+
$$
\vskip0.2cm
$$
+\tr^2\Bigl(\h Q({\h\dl}+e^{\be\h\ep})\1\Bigr)-\tr
\Bigl(\h Q({\h\dl}+e^{\be\h\ep})\1 \h Q({\h\dl}+
                                  e^{\be\h\ep})\1\Bigr),
$$
etc.;

\vskip0.3cm
\leftline{b) in the $XY$-case:}

$$
\lim_{\al\to0}
    \frac{(d/d\al)\Det^{1/2} A_\al}{\Det^{1/2} A_\al}=
-\frac12\,\tr\!\left(\frac{{\w g}\,\Bigl({\h Q}
        \otimes\sigma^3\Bigr)\,{\w g}^{-1}}
  {{\w I}+\exp\big(\!-\be{\h E}\otimes\sigma^3\big)}\right),
$$
$$
\eqno(6.27)
$$
$$
\lim_{\al\to0}
   \frac{(d^2/d\al^2)\Det^{1/2} A_\al}{\Det^{1/2} A_\al}=
\frac12\,\tr\!\left(\frac{{\w g}\,\Bigl({\h Q}
        \otimes I \Bigr)\,{\w g}^{-1}}
  {{\w I}+\exp\big(\!-\be{\h E}\otimes\sigma^3\big)}\right)
$$
\vskip0.2cm
$$
+\frac14\,\tr^2 \left(\frac{{\w g}\,\Bigl({\h Q}
        \otimes\sigma^3\Bigr)\,{\w g}^{-1}}
{{\w I}+\exp\big(\!-\be{\h E}\otimes\sigma^3\big)}\right)-
\frac12\,\tr\left(\frac{{\w g}\,\Bigl({\h Q}
        \otimes\sigma^3\Bigr)\,{\w g}^{-1}}
{{\w I}+\exp\big(\!-\be{\h E}\otimes\sigma^3\big)}\,\,
\frac{{\w g}\,\Bigl({\h Q}
        \otimes\sigma^3\Bigr)\,{\w g}^{-1}}
{{\w I}+\exp\big(\!-\be{\h E}\otimes\sigma^3\big)}
\right),
$$
etc.

Equations (6.26), (6.27) are obtained with the help of
(6.24), (6.25), i.e., from the integral representations
of the type (6.1). As an additional check of consistency 
between (6.24), (6.25), on the one hand, and the results 
given by finite-dimensional determinants deduced in 
Subsection 6.1, on the other, one can verify that appropriate 
differentiations of $\det ({\h\dl}+{\h M}_F(\al))$ (6.6) 
(with ${\h M}_F (\al)$ (6.7b)) over $\al$ also result in 
(6.26). Similarly, differentiations of $\det^{1/2}({\w I}
+{\w M}_F(\al))$ (6.12), where ${\w M}_F(\al)$ is given by 
(6.19c), also lead us to (6.27).

As an other example, let us calculate, in the thermodynamic 
limit, the derivatives $(G^\pm_F)_\al^{\,\prime}$ and 
$(G^\pm_F)_\al^{\,\prime\prime}$ at $\al=0$ using, for
a comparison, (6.12) and (6.17). To proceed, we adopt the 
following definitions:
$$
C_q \equiv 1-\cos\theta_q\,\th\frac{\be E_q}{2}, \quad
S_q \equiv \sin\theta_q\,\th\frac{\be E_q}{2}.
\eqno(6.28)
$$
It is straightforward to establish the following
relation:
$$
\frac{d}{d\al}\,G^\pm_F\,{\Bigg |}_{\al=0} =
\tr \Bigl({\h K}_F^{\,\prime} (0)\Bigr)
$$
$$
=\frac 12 \left[m+\tr\Bigl({\w M}_F^{\,\prime} (0)\Bigr)\right]
=\frac m2\,\int\limits^{\pi}_{-\pi}C_q\,\frac{dq}{2\pi},
\eqno(6.29)
$$
where $C_q$ is given by (6.28). Further,
$$
\frac{d^2}{d\al^2}\,G^\pm_F\,{\Bigg |}_{\al=0} =
\tr^2 \Bigl({\h K}_F^{\,\prime} (0)\Bigr)
+\tr \Bigl({\h K}_F^{\,\prime\prime} (0)\Bigr)
-\tr \Bigl({\h K}_F^{\,\prime} (0)\,
                      {\h K}_F^{\,\prime} (0)\Bigr)
\eqno(6.30a)
$$
$$
=\frac 14 \left[m+\tr\Bigl({\w M}_F^{\,\prime} (0)\Bigr)
\right]^2 + \frac 12 \tr\Bigl({\w M}_F^{\,\prime\prime} (0)\Bigr)
-\frac 12 \tr\Bigl({\w M}_F^{\,\prime} (0)\,
                        {\w M}_F^{\,\prime} (0)\Bigr).
\eqno(6.30b)
$$
We take into account that $\tr\Bigl({\w M}_F^{\,\prime\prime} 
(0)\Bigr)=m$, $\,\tr \Bigl({\h K}_F^{\,\prime\prime} (0)\Bigr)=
\tr \Bigl({\h K}_F^{\,\prime} (0)\Bigr)$, and
$$
-\frac 12 \tr\Bigl({\w M}_F^{\,\prime} (0)
                        {\w M}_F^{\,\prime} (0)\Bigr)
= -\frac m2\,\int\limits^{\pi}_{-\pi} 
       \cos\theta_q\,\th\frac{\be E_q}{2}\,\frac{dq}{2\pi}
$$
$$
+\frac14 \int\limits^{\pi}_{-\pi}\int\limits^{\pi}_{-\pi}
Q_{pq} Q_{qp} (S_p S_q - C_p C_q)\,dp\,dq,
$$
where $C_q$ and $S_q$ are given by (6.28). Then, direct use of 
explicit form of ${\h K}_F^{\,\prime} (0)$ to express
$\tr \Bigl({\h K}_F^{\,\prime} (0)\,
{\h K}_F^{\,\prime} (0)\Bigr)$ in (6.30a) demonstrates us
that (6.30a) and (6.30b) coincide. 

To conclude the section, let us use the obtained formulas 
(6.29), (6.30) to calculate the correlators $\l\si_{m}^z\r$ 
and $\l\si_{m+1}^z\si_{1}^z\r$ in the thermodynamic limit. 
In this case, the contributions $G^+_B Z^+_B$ and $G^-_B Z^-_B$
in (2.9) cancel and only derivatives of $G^\pm_F$ are important.
We obtain for $\l Q^n(m)\r$:
$$
\l Q(m)\r = \frac{m}{4\pi} \int\limits^{\pi}_{-\pi}C_q dq,
\quad
\l Q^2(m)\r = \frac{m}{4\pi} \int\limits^{\pi}_{-\pi}C_q dq +
\biggl(\frac{m}{4\pi} \int\limits^{\pi}_{-\pi}C_q dq\biggr)^2
$$
$$
+\frac14 \int\limits^{\pi}_{-\pi}\int\limits^{\pi}_{-\pi}
Q_{pq} Q_{qp} (S_p S_q - C_p C_q)\,dp\,dq,
\eqno(6.31)
$$
where $C_q, S_q$ are given by (6.28). The usage of (6.20) 
instead of (6.12) also leads to (6.31). The relations 
obtained lead to the correct answers for the correlation 
functions \cite{col}, \cite{iz1}, \cite{iz2}:
$$
\l\sigma^z_m\r = (1/2\pi)\int\limits^{\pi}_{-\pi}
\cos\theta_q\,\th\frac{\be E_q}{2} dq,
\eqno(6.32)
$$
where the definitions
$$
\l\si_{m}^z\r=1-2\CD_1\l Q(m)\r
$$
and $\CD_1 f(m)=f(m)-f(m-1)$ are used. We obtain
further:
$$
\l\sigma^z_{m+1}\sigma^z_1\r - \l\sigma^z_m\r^2 =
(1/4\pi^2) \int\limits^{\pi}_{-\pi}\int\limits^{\pi}_{-\pi}
\cos m(p-q) (S_p S_q - C_p C_q)\,dp\,dq\,,
\eqno(6.33)
$$
where the definitions
$$
\l\si_{m+1}^z\si_{1}^z\r=2\CD_2\l Q^2(m)\r
+2\si_z-1
$$
and $D_2f(m)=f(m+1)-2f(m)+f(m-1)$ are used.

In the $XX$-case we obtain, say, from (6.26): 
$$
\l Q{(m)}\r=\frac m{2\pi}\,\int\limits_{-\pi}^\pi 
                (1+e^{\be\ep_q})\1 dq,
\eqno(6.34)
$$
$$
\l Q^2(m)\r=\frac m{2\pi}\,\int\limits_{-\pi}^\pi 
             (1+e^{\be\ep_q})\1 dq
+\left( \frac{m}{2\pi} \int\limits_{-\pi}^\pi 
           (1+e^{\be\ep_q})\1 dq \right)^2-
$$
$$
-  \int\limits^{\pi}_{-\pi}\int\limits^{\pi}_{-\pi}
           Q_{pq}^2 (1+e^{\be\ep_p})\1
               (1+\ep^{\be\ep_q})\1 dp\,dq.
\eqno(6.35)
$$
One obtains from (6.34) in the thermodynamic limit:
$$
\si^z\equiv\l\si_{m}^z\r=1-\frac1\pi \int\limits^\pi_{-\pi}
\frac {dq}{1+e^{\be\ep_q}},
\eqno(6.36)
$$
The result (6.36) agrees with the magnetization 
$M_z=-\cd F/\cd h$, which is calculated by means
of $F$ (5.12). We obtain from (6.35) ($m>0$):
$$
\l\si_{m+1}^z\si_{1}^z\r=(\si^z)^2 -\frac1{\pi^2}
\left|\int\limits^\pi_{-\pi} \!\frac{e^{imq}}{1+e^{\be\ep_q}}
\,\,dq\right|^2.
\eqno(6.37)
$$

Equations (6.32), (6.33) reproduce (6.36), (6.37) in the limit 
$\ga\to 0$, i.e., at $\theta_q\to 0$ in $C_q$, $S_q$. The 
answers obtained (6.32), (6.33) agree with the results of 
\cite{col}, \cite{iz1}, \cite{iz2}, \cite{ml}. Therefore they 
witness in favour of self-consistency of the functional-integral 
representations, given by (3.12) (to be used in (2.9)) with the 
corresponding boundary conditions, which appear when calculating,
in the way inspired by \cite{alv1}, the generating functional 
$G(\al, m)$ (2.1) for the $XY$ and $XX$ Heisenberg cyclic chains. 
The boundary conditions in the imaginary time which follow 
(3.12) are given by (3.8) (for $G^\pm_F Z^\pm_F$) and by (3.13) 
(for $G^\pm_B Z^\pm_B$). Besides, our regularizations resulting in 
the general expressions (6.6), (6.7), (6.12), (6.19) look
satisfactory since final results agree with the results obtained 
by other methods.

\section{ Discussion}

The generating functionals of static correlators of third 
components of local spins for the $XY$ and $XX$ Heisenberg spin 
{\it 1/2} chains are calculated in the present paper. The 
calculation is carried out by means of the functional 
integration over trajectories with ``automorphic'' dependence 
on the imaginary time. The results are obtained in the form of 
determinants of the matrix operators which are regularized 
by means of zeta-regularization. It is demonstrated that, 
from a practical standpoint (i.e., if only differentiations 
over the parameter $\al$ are needed), the formula for the 
first derivative over $\al$ of the generating functional can be 
obtained without regularization of the Mellin integral at 
the lower bound. Various special limits are considered for 
the formulas obtained which demonstrate reductions of the 
$XY$-case to the known relations of the $XX$-case. 

The given paper continues \cite{mal1} where a 
method has been proposed which allows to calculate 
vacuum average of an exponential of quadratic operator as a 
functional integral over ``automorphic'' trajectories. 
The present paper is close to \cite{alv1} where the 
functional integration with ``automorphic'' boundary 
conditions is used for a calculation of certain differential
geometric indices. The present paper is also 
close to \cite{fed}, where the partition functions 
of spin {\it 1/2} and spin {\it 1} chain models have been also 
obtained in the form of path integrals over variables subjected 
to ``automorphic'' boundary conditions. The distinction of 
the present paper from \cite{alv1}, \cite{fed} 
consists in the fact the ``automorphic''boundary condition 
in the imaginary time appears only for a part of sites. The 
method proposed in \cite{mal1} is considered above for the system 
which is equivalent to quasi-free fermions (i.e., the 
corresponding Hamiltonian is diagonalized by the Bogoliubov 
transformation). Approach presented provides further 
development of the technical finds discussed in \cite{alv1}, 
\cite{fed}, \cite{gr}, and it seemingly merits attention 
since can be used further, for instance, for the $XX$ Heisenberg 
model with translationally inhomogeneous boundary conditions. 
In general, the functional integral considered merits attention 
since it can be useful, as a technical method, for other models 
where it is also necessary to calculate vacuum 
averages of exponentials of quadratic 
operators of the type of $\exp(\al Q(m))$.

\section*{Acknowledgement}

The research described has been supported in part by RFBR,
Projects No. 01-01-01045, 04-01-00825

\vskip 1.0cm
\noindent
{\it NOTE ADDED IN PROOF:}

\noindent
Equation (6.18) demonstrates, in fact, that the square root in
$G^\pm_F$ (6.12) can be calculated at $\gamma =0$ thus leading
to an expression in the form of the determinant of the matrix 
of the size $M\times M$. The same is true for $\gamma\ne 0$
also: calculation of the square root in (6.12) at $\gamma\ne 0$
leads to $G^\pm_F$ just in the form (6.17). The corresponding details
should be presented elsewhere.

\end{document}